\documentclass [aps,pra,amssymb,amsmath,showpacs,twocolumn]{revtex4-1}
\usepackage[latin9]{inputenc}
\usepackage{amsmath}
\usepackage{amssymb}
\usepackage{graphicx}
\usepackage{verbatim}
\usepackage{bm}
\usepackage{color}

\usepackage{graphicx,epsfig,amsfonts,amssymb}
\usepackage{bm}
\usepackage{times}
\usepackage{lipsum}
\usepackage{verbatim}

\newcommand{\ra}{\rangle}
\newcommand{\rar}{\rightarrow}

\usepackage{hyperref}
\hypersetup{
    colorlinks,
    citecolor=blue,
    filecolor=blue,
    linkcolor=blue,
    urlcolor=blue
}

\newcommand{\beg}{\begin{equation}}
\newcommand{\en}{\end{equation}}

\newcommand{\eref}[1]{Eq.~(\ref{#1})}
\newcommand{\re}[1]{(\ref{#1})}

\newcommand{\eps}{\varepsilon}

\newcommand{\be}{\begin{eqnarray}}
\newcommand{\ee}{\end{eqnarray}}

\newcommand{\bs}{\begin{equation}\begin{split}}
\newcommand{\es}{\end{split}\end{equation}}

\date{\today}

\begin{document}
\title{Integrable   time-dependent quantum Hamiltonians}
\author{Nikolai A. Sinitsyn$^a$, Emil A. Yuzbashyan$^b$, Vladimir Y. Chernyak$^c$, Aniket Patra$^{a,b}$, and Chen Sun$^{a,d}$}
\affiliation{$^a$Theoretical Division, Los Alamos National Laboratory, Los Alamos, NM 87545,  USA}
\affiliation{$^b$Center for Materials Theory, Department of Physics and Astronomy, Rutgers University, Piscataway, NJ 08854, USA}
\affiliation{$^c$Department of Chemistry and Department of Mathematics, Wayne State University, 5101 Cass Ave, Detroit, Michigan 48202, USA}
\affiliation{$^d$ Department of Physics, Texas A\&M University, TX 77840,  USA}


\begin{abstract}
We formulate a set of conditions under which dynamics of a time-dependent quantum Hamiltonian are integrable.
The main requirement is the existence of a nonabelian gauge field with zero curvature    in the space of system parameters. Known solvable multistate Landau-Zener models satisfy these conditions.    Our method  provides  a strategy to  incorporate time-dependence  into  various quantum integrable models, so that the resulting non-stationary Schr\"odinger equation is  exactly solvable. We also validate some prior  conjectures, including  the solution of the driven generalized Tavis-Cummings model.
\end{abstract}

\maketitle

 Quantum coherent dynamics controlled by strong time-dependent fields can be realized and explored nowadays in systems of considerable complexity
\cite{kinoshita2006,monaco,ligner,weiler,widera,Gring:2012,Trotzky:2012,shimano2013}. Time-dependent parameters play a critical role in NMR \cite{NMR-LZ}, quantum information processing \cite{dot-lz-exp1,nichol,zhou,gaudreau,LZ-metrology,forster,riberio,wubs,Wernsdorfer,annealing-LZ,q-phaset}, molecular dynamics \cite{book-LZ,child,nikitin} and  cold atom experiments \cite{regal,zwierlein,atomic}. On the theory side, quantum
dynamics of  time-dependent many-body Hamiltonians, especially their exact analytical description, present  considerable challenges. In contrast,  exact solutions of significant relevance to experiment inform our understanding of stationary states, e.g., Bethe's Ansatz solution of  paradigmatic models \cite{gaudin,bethe2,sutherland}. Nontrivial exact results have been also obtained for quantum quenches, such as the Generalized Gibbs Ensemble description of the dynamics of the spin-1/2 Heisenberg chain \cite{ilievski2} and quantum quench phase diagrams of BCS superconductors \cite{quench1}. Such methods, unfortunately, do not apply to a Hamiltonian with continuous time-dependence. 

In this letter, we propose an approach for solving the non-stationary Shr\"odinger equation exactly for a certain class of time-dependent Hamiltonians. This approach allows us to make parameters of  a quantum integrable model, e.g., the BCS and generalized Tavis-Cummings Hamiltonians, vary in time in such a way that resulting dynamics  are exactly solvable. Here we focus primarily on the scattering problem, i.e. on determining the time-evolution over a specific time interval.

Important examples of driven systems are matrix Hamiltonians linear in time, $ H(t) =  A + Bt$, where $A$ and $B$ are time-independent Hermitian $N\times N$ matrices. The problem of finding the scattering matrix that relates the state of the system at $t=+\infty$ to that at $t=-\infty$ is called then the multistate Landau-Zener problem. The $2\times2$ problem was solved by Landau, Zener, Majorana and 
St\"uckelberg in 1932 \cite{landau,zener,Majorana,stuckelberg}. For $N\ge3$  the solution is known only for special choices of $A$ and $B$. Earliest examples include Demkov-Osherov \cite{do,be}, bow-tie \cite{bow-tie}, generalized bow-tie  \cite{gbow-tie1,gbow-tie}, composite \cite{multiparticle}, and infinite chain
\cite{lz-chain} models. In a more recent work \cite{yuzbashyan-LZ}, it was shown that
nontrivial solvable models belong to families of mutually commuting Hamiltonians linear or quadratic in $t$. It was therefore conjectured that quantum integrability understood as the existence of nontrivial time-dependent commuting partners \cite{shastry,haile1,haile,shastry1} is a necessary condition for the multistate Landau-Zener solvability. In a parallel development, methods to solve and search for new models  were discovered  \cite{six-LZ,quest-LZ,constraints}  and since then the number of such models has   grown rapidly~\cite{four-LZ,DTCM,DTCM1,chen-largeLZ}.  

Our approach provides a unified framework to derive  exact solutions for all these models and supports the conjecture made in \cite{yuzbashyan-LZ}. Below, we  first formulate our approach and then discuss various many-body and matrix models that fit into it. To illustrate our technique, we solve  the scattering problem for two nontrivial models  -- the generalized Tavis-Cummings Hamiltonian with a linear drive and a new 4-state Hamiltonian linear in $t$. We conclude  with several general observations and an outline of the idea of the solution for arbitrary $t$.

Consider a Hamiltonian $\hat H(t, \vec x)$ that, in addition to time, depends on $M$ real parameters $(x^1,\dots, x^M)=\vec x$. For example, in the multistate Landau-Zener problem these can be certain matrix elements of $A$ and $B$. The main idea is to embed the non-stationary Shr\"odinger equation for $\hat H(t, \vec x)$
into a set of multi-time  Shr\"odinger equations
\begin{equation}
\label{system1}
 i \partial_{j} \Psi(\bm{x}) = \hat{H}_{j} \Psi(\bm{x}), \; \phantom{\sum} j = 0, 1, \ldots, M,
\end{equation}
where $\bm x=(t,\vec x)$,  $\partial_j\equiv \partial/\partial x^j$, $x^0=t$,  $\hat H_0\equiv \hat H(t, \vec x)$, and Hamiltonians $\hat H_j$ are independent. In other words, the first equation ($j=0$) is 
the non-stationary Shr\"odinger equation, while the rest are auxiliary Shr\"odinger equations that help us solve it exactly. Taking the derivative of \eref{system1} with respect to $x^k$, we derive consistency conditions 
\begin{eqnarray}
\label{system11}
 \partial_{j} \hat{H}_{k} - \partial_{k} H_{j} - i [\hat{H}_{k}, \hat{H}_{j}] = 0, \quad k, j = 0, \ldots, M.
\end{eqnarray}
These conditions are sufficient and necessary for system \re{system1} to possess a joint solution for any initial condition \cite{petrat,deform-book}. We may view them as a generalization of the notion of integrals of motion for time-dependent quantum Hamiltonians.

A formal solution of \eref{system1} along a path  in the space of real parameters $\bm x$ that starts at a reference point $\bm{x_{0}}$
 is  an ordered exponential
\begin{eqnarray}
 \label{time-ord}
\Psi(\bm{x}) = T \exp\left(-i \int_{{\cal P}} \hat{H}_{j} dx^{j}\right) \Psi(\bm{x_{0}}),  
\end{eqnarray}
 where we assume summation over repeated  indices.  Treating  Hamiltonians $\hat H_j$
as matrix components of a nonabelian gauge field ${\cal A}(\bm{x})$,  ${\cal A}_{j} = -i\hat{H}_{j}$, we interpret  Eq.~(\ref{system11}) as the zero curvature condition ${\cal F}_{jk}\equiv \partial_j{\cal A}_{k}-\partial_k{\cal A}_{j}-[{\cal A}_{j},{\cal A}_{k}]=0$, so that
$\Psi(\bm{x})$ in Eq.~(\ref{time-ord}) is independent of  the integration path ${\cal P}$ as long as its endpoints are fixed. Similar zero curvature integrability condition is also well known in soliton physics~\cite{faddeev-book}. It is precisely this freedom to choose a suitable path  that enables us to explicitly solve the scattering problem. 

Further, consider a path ${\cal P}_\tau$ parameterized by a variable $\tau$
\begin{eqnarray}
 \label{timec1} {\cal P}_\tau:\quad x^{j}(\tau) = v^{j} \tau + x_{0}^{j}, \quad j=0,\ldots, M,
\end{eqnarray}
where $v^{j}$ and $x_{0}^{j}$ are constants.
The state vector $\Psi(\tau) = \Psi(\bm{x}(\tau))$ along this path satisfies 
\begin{eqnarray}
\label{dynx1} i\frac{d\Psi(\tau)}{d\tau} &=& \hat{h}(\tau) \Psi(\tau), \\
\label{comb1} \hat{h}(\tau) &=& \sum_{j} v^{j}\hat H_{j}(\bm{x}(\tau)).
 \end{eqnarray}
 Solutions of \eref{dynx1} follow from those of \eref{system1}. Therefore,  $\hat h(\tau)$ -- an arbitrary linear combination of $\hat H_j$ -- is also a solvable time-dependent model just like a linear combination of integrals of motion of a time-independent model is also an integral. Note however that the coefficients $v^j$ of this linear combination dictate the time-dependence of $\hat h(\tau)$.
 
 An important observation is that complex looking Eq.~(\ref{system1})  simplifies considerably when  the matrix elements of the Hamiltonians are real. Then, the real and imaginary parts of \eref{system11} yield two separate conditions
 \begin{eqnarray}
\label{commute-p1}
\left[\hat{H}_{j},\hat{H}_{k} \right] &=& 0,  \\
\label{commute-p2}
\partial_{j} \hat{H}_{k} - \partial_{k} \hat{H}_{j} = 0, &&\,\, j,k=0, 1,\ldots, M.
\end{eqnarray}
These equations
suggest a strategy for identifying solvable time-dependent models.
 First, we note that \eref{commute-p1} is to be supplemented with a notion of a nontrivial commuting partner  that weeds out trivial
 partners    (e.g., projectors onto the eigenstates of $\hat H$). One way is to restrict the parameter (time) dependence of $\hat H_j$ to be linear or, more generally, polynomial in $t$. This leads to a systematic classification and explicit construction of commuting families of parameter-dependent matrix Hamiltonians \cite{haile,haile1,shastry,yuzbashyan-LZ}, which are interesting candidates for our approach.
 More generally, any quantum integrable model that contains two or more real parameters is a potential candidate. Such models have an extensive number of integrals of motion that satisfy \eref{commute-p1}. If no initial subset of integrals satisfies \eref{commute-p2}, we attempt to redefine them by taking  various combinations and similarly redefine the parameters to make \eref{commute-p2} work for at least $M=1$. Note that once we declare one of the variables $x^j$ to be the physical time,  commuting partners $\hat H_j$ cease to be integrals of motion.

For example, take the generalized Tavis-Cummings model
\be
\hat{H}_\mathrm{TC}= \sum_{j=1}^{N_s} \varepsilon_j \hat{s}_j^z -\omega \hat{a}^{\dagger} \hat{a} + g \!\sum_{j=1}^{N_s}(\hat{a}^{\dagger} \hat{s}^-_j + \hat{a} \hat{s}^{\dagger}_j ),
\label{TC}
\ee
where $\hat{a}$ is the boson annihilation operator and $\hat{s}_j^z$, $\hat{s}_j^{\pm}$ are spin-1/2 operators. Its commuting partners are \cite{dicke}
\be
\hat{H}_j\!  =\!  (\varepsilon_j \! +\! \omega) \hat{s}_j^z \! +\! g(\hat{a}^{\dagger} \hat{s}_j^-\! \!  + \! \hat{a} \hat{s}_j^{\dagger} ) \! + \! 2g^2\!  \sum_{k \ne j} \frac{\hat{\bf s}_j \cdot   \hat{\bf s}_k}{\varepsilon_j -\varepsilon_k}.
\label{gaudin1}
\ee
Equations~(\ref{commute-p1},\ref{commute-p2}) hold with $M=N_s$, $\hat H_0 = \hat{H}_\mathrm{TC}$, $\bm{x}=(\omega, \eps_1,\dots,\eps_{N_s})$. Another
example is the BCS Hamiltonian. In terms of Anderson pseudospin-1/2 operators it reads
\beg
\hat H_\mathrm{BCS}=\sum_{j=1}^{N_s} 2\eps_j \hat s_j^z - \frac{1}{2B}\sum_{j,k} \hat s_j^+ \hat s_k^-,
\en
where $(2B)^{-1}$ stands for the BCS coupling constant. Its commuting partners are Gaudin magnets \cite{gaudin,integ}
\beg
\hat H_j=2B\hat s_j^z-\sum_{k\ne j}\frac{\hat {\bf s}_j\cdot \hat {\bf s}_k}{\eps_j-\eps_k}.
\label{gaudin7}
\en
Now $\hat H_0 = \hat{H}_\mathrm{BCS}$ and $x^0=B$. Thus,
the generalized Tavis-Cummings model with a linear sweep of the bosonic frequency, $\omega=t$, and the BCS Hamiltonian with coupling $\propto 1/t$ both fit into our construction. Similarly, using the commuting partners derived in  \cite{yuzbashyan-LZ}, we verified  that   the Demkov-Osherov, bow-tie,  and generalized bow-tie, as well as Landau-Zener-Coulomb models \cite{sinitsyn-14pra,LZC,LZC1,LZC2} fit into our construction. 

A key point of this letter is that   zero curvature condition (\ref{system1}) leads to an explicit exact solution of the scattering problem. Consider, e.g., the multistate Landau-Zener model $\hat H(t,\vec x)=\hat A(\vec x)+t\hat B(\vec x)$ for which we need to determine the matrix  of  transition probabilities $P$ with elements ${P}_{nn'}\equiv {P}_{n' \rar n} \equiv  |S_{nn'}|^2$. Here $S$ is the scattering matrix between eigenstates at $t=-\infty$ and $t=+\infty$ at some fixed values of the parameters, $\vec x=\vec c$ \cite{be}. As discussed above, we are free to choose any path in the space $\bm{x}=(t, \vec x)$ that connects the points $(-\infty, \vec c)$ and $(+\infty, \vec c)$. It is convenient to choose a path ${\cal P}_\infty$, such that $|\bm{x}|$ is  always large and the time-evolution is adiabatic   everywhere, except the neighborhood of isolated points, where scattering takes place. The corresponding scattering problem is typically simple thanks to large $|\bm{x}|$, e.g., it reduces to a $2\times2$ Landau-Zener problem in the two nontrivial examples we consider below. In general, Eqs.~(\ref{commute-p1},\ref{commute-p2}) enable one to construct a multidimensional version of WKB    with simple scattering matrices   connecting  adiabatic (WKB)  solutions in different adiabatic regions \cite{suppl}.

\begin{figure}
\vskip -0.1in
\includegraphics[width=7.cm]{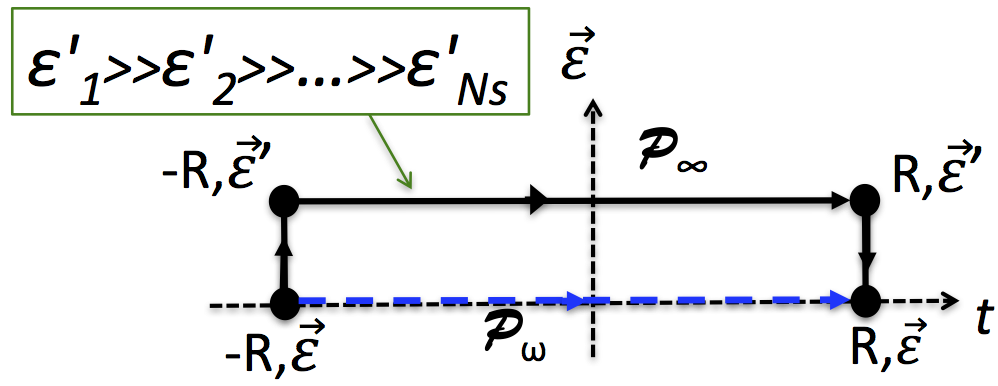}
\centering
\caption{(color online) Paths in the space of  parameters $\eps_j$ and time used to evaluate  transition probabilities
for the driven generalized Tavis-Cummings  model (\ref{TC}), $\omega=t$. On ${\cal P}_\omega$, $\omega$ changes from $-R$ to $+R$, all $\eps_j$ are fixed. Since this model is a part of a commuting family of $\hat H_j$ that satisfies the zero curvature condition, we can deform ${\cal P}_\omega$ into a new path ${\cal P}_\infty$ without modifying the scattering matrix.}
\label{contour2}
\end{figure}

 Our first example is the Tavis-Cummings model \re{TC} with linear drive, $\omega=t$. Let $\eps_1>\eps_2>\dots>\eps_{N_s}$. We are interested in the evolution  along the path ${\cal P}_{\omega}$ shown in
Fig.~\ref{contour2}. On this path $\eps_j=\mbox{const}$, while $\omega$ changes from $-R$ to $R$. At the end we take the limit $R\to\infty$. This scattering 
problem was solved in \cite{DTCM} under the assumption that $\eps_j$ are well separated, i.e. $\eps_1\gg\eps_2\gg\dots \gg\eps_{N_s}$. It was further conjectured in \cite{DTCM} that this is the general solution. We are now in the position to prove this conjecture. To do so, consider the path ${\cal P}_\infty$ in Fig.~\ref{contour2} that has the same endpoints as ${\cal P}_\omega$. On the first vertical leg of ${\cal P}_\infty$, $\eps_j$ evolve,  keeping the ordering of $\eps_j$,  until the condition $\eps_1\gg\eps_2\gg\dots \gg\eps_{N_s}$ is met. On the second vertical leg, they evolve back to their initial values. Since $|\omega|$ is large and $\eps_j$ are distinct, this evolution is purely adiabatic and does not affect the transition probabilities. On the horizontal leg of ${\cal P}_\infty$ the problem is precisely the one solved in \cite{DTCM}. This proves the above conjecture.

In our second example, we take a previously solved $4\times 4$ multistate Landau-Zener problem    \cite{four-LZ,constraints}  and derive from it a new,  more general
Hamiltonian by the prescription outlined below \eref{comb1}. We then proceed to determine the transition probabilities for this new model. Let
\be
{H}(t,e) = \left(
\begin{array}{cccc}
b_1t+e & 0 & g &-\gamma \\
0 & -b_1t +e & \gamma & g \\
g& \gamma & b_2 t & 0\\
-\gamma & g &0 &-b_2t
\end{array}
\right),
\label{ham1}
\ee
where $b_1$, $b_2$, $e$, $g$ and $\gamma$ are constants.  To determine if   this Hamiltonian fits into our approach, 
we first search for  a nontrivial commuting partner $H_1$ linear in $t$.   This reduces to a set of linear algebraic equations for parameters of ${H}_1$  \cite{haile}.  We find three  linearly independent commuting operators. Two of them are trivial -- the unit matrix and  ${H}$ itself. Therefore, there is a single nontrivial  commuting partner, which we determine explicitly.
When both $H_0\equiv{H}$ and ${H}_1$ are linear in $t$ ,  \eref{commute-p1} implies that their time-dependent parts are diagonal in the same  basis. So,  to satisfy (\ref{commute-p2}), the parameter $x^1$ must be constructed from  diagonal time-independent elements of ${H}$. A natural candidate   is  $x^1=e$. Searching then for ${H}_1$ that satisfies (\ref{commute-p2}) in the form of a linear combination of the three commuting operators, we find
\be
{H}_1(t,e) \!= \! \left(\!\!
\begin{array}{cccc}
t \!+\!\frac{b_1e}{b_1^2-b_2^2} & 0 & \frac{g}{b_1-b_2} &\frac{-\gamma}{b_1+b_2} \\
0 & t \!-\! \frac{b_1e}{b_1^2-b_2^2}  & \frac{-\gamma}{b_1+b_2} & \frac{-g}{b_1-b_2} \\
\frac{g}{b_1-b_2}& \frac{-\gamma}{b_1+b_2} &- \frac{b_2e}{b_1^2-b_2^2}& 0\\
\frac{-\gamma}{b_1+b_2} & \frac{-g}{b_1-b_2} &0 &\frac{b_2e}{b_1^2-b_2^2}
\end{array}
\!\!\right)\!\!.
\label{ham2}
\ee
Let the evolution path be
\be
{\cal P}_\tau: \,\, t = \tau, \,\, e = v\tau + e_0,
\label{cont1}
\ee
with constant $v$ and $e_0$. The  Hamiltonian (\ref{dynx1}) for ${\cal P}_\tau$ is
\be
\label{heff1}
&& h(\tau)\!  =\!\! \left(\!\! 
\begin{array}{cccc}
\beta_1\tau+e_1& 0 & g(1+x) &-\gamma(1+ y) \\
0 & \beta_2 \tau+e_2  &\gamma (1-y) & g(1-x)\\
g(1+x)& \gamma (1-y) &\beta_3 \tau + e_3& 0\\
-\gamma (1+y) & g(1-x) &0 & \beta_4 \tau +e_4
\end{array}\!\! 
\right)\!\! , \\
\nonumber && x=\frac{v}{b_1-b_2}, \,\, y=\frac{v}{b_1+b_2}, \,\, \beta_{1,2}= 2v \pm b_1(1+xy),\, \beta_{3,4}= \\
\nonumber && \pm b_2(1 - xy), \, e_{1,2}= e_0 (1\pm b_1 xy/v), \,e_{3,4}=\mp e_0 b_2 xy /v.
\ee
This is a new, previously unsolved   model more general than (\ref{ham1}), e.g., all  couplings (off-diagonal matrix elements) in (\ref{heff1}) are distinct.  We proceed to solve it with our method.

\begin{figure}
\includegraphics[width=8.75 cm]{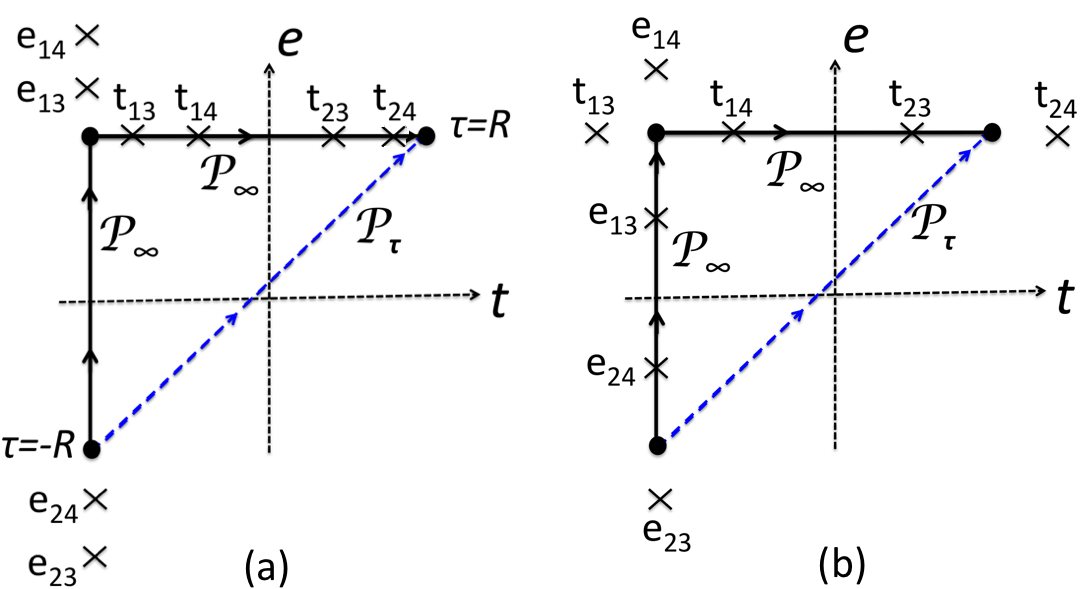}
\centering
\caption{(color online) Paths  in the space $(t,e)$ for evaluating  transition probabilities
for the  model (\ref{heff1}). On ${\cal P}_\tau$, $\tau$ changes from $-R$ to $+R$; all other parameters are fixed. We  deform ${\cal P}_\tau$ into ${\cal P}_\infty$ without affecting the scattering matrix.
 Points $e_{ij}$ and $t_{ij}$ marked with crosses indicate nonadiabatic  Landau-Zener  transitions between levels $i$ and $j$ for (a) $v<b_1-b_2$ and (b) $b_1+b_2>v>b_1-b_2$.}
\label{contour1}
\end{figure}

Let $b_1>b_2>0$ and $v>0$. We are interested in the evolution matrix for $h(\tau)$ along the path ${\cal P}_\tau$  from $\tau = -R$ to $\tau = R$, see Fig.~\ref{contour1}(a), in the limit $R \rar \infty$.  Because ${H}_{0}(t, e)$ and ${H}_1(t, e)$ satisfy the zero curvature condition, the evolution matrix is  the same as that for the path ${\cal P}_{\infty}$. The latter  has two pieces. In the vertical one, we set $t=-R$ and vary $e$ from $-vR +e_0$ to $vR+e_0$. In the horizontal piece, we fix $e=vR+e_0$ and vary $t$ from $-R$ to $R$. According to \eref{time-ord}, only  ${H}_1$ contributes on the first piece  and only ${H}_{0}$  on the second. Along ${\cal P}_{\infty}$,  diagonal matrix elements of  $H_0$ and $H_1$ (diabatic levels) are large compared to the couplings. Therefore,  the levels   are   well separated, except on disjoined small segments of  ${\cal P}_{\infty}$ near points where a pair of the diagonal elements  is degenerate. These segments connect adiabatic parts of ${\cal P}_{\infty}$ where the adiabatic approximation  is exact in the limit $R\to\infty$. Let us write the state of the system as $\Psi(t,e)=\sum_k a_k|k\rangle$, where $|k\rangle$ are the eigenstates of the diagonal parts of  $H_0$ and $H_1$ (diabatic eigenstates). Diabatic and adiabatic (instantaneous) eigenstates coincide in adiabatic parts of ${\cal P}_{\infty}$  when $R\to\infty$.
 In the adiabatic approximation, absolute values of $a_k$ remain the same, while their phases evolve with $t$ and $e$. 
 
In the vicinity of degeneracy points, two levels come close and transitions between them become locally possible. The other two  levels, however, remain far remote and do not affect these nonadiabatic transitions. Suppose $v < b_1-b_2$. For this case, we mark the points of diabatic level crossings with crosses in Fig.~\ref{contour1}(a). Along ${{\cal P}}_{\infty}$,  adiabatic approximation brakes near  four points that all have $e=vR+e_0$ and
$$
t_{13/24}=\mp \frac{vR+e_0}{b_1-b_2}, \quad t_{14/23} =\mp \frac{vR+e_0}{b_1+b_2}.
$$
The distances between these points are $\propto R$, which means that regions  of  pairwise nonadiabatic transitions along ${{\cal P}}_{\infty}$ are well apart. Consider, e.g., the evolution of the amplitudes $a_1$ and $a_3$   near  $t_{13}$ that is governed by $H_0$. Writing $t=t' + t_{13}$ and disregarding the other two levels, we find
\be
i\frac{d a_1}{d t'} = b_1 t' a_1+ g a_3, \quad i \frac{da_3}{dt'} = b_2t' a_3 +g a_1,
\label{ev13}
\ee
which is a $2\times2$ Landau-Zener problem, whose scattering matrix is known explicitly  \cite{landau,zener,Majorana,stuckelberg}.
Since the other two levels do not experience nonadiabatic transitions here, they produce only diagonal unit entries in the scattering matrix for evolution through $t_{13}$. The total evolution matrix $S$ for the path ${\cal P}_{\infty}$ factorizes into an ordered product of such pairwise scattering matrices ${S}^{ab}$, where $a, b$  label  states experiencing  nonadiabatic transitions and diagonal matrices $U^{\alpha,\beta}$ describe adiabatic evolution between points $\alpha$ and  $\beta$ on this path, i.e.
$$
{S} = {U}^{R,t_{24}} {S}^{24} {U}^{t_{24},t_{23}} {S}^{23}  {U}^{t_{23},t_{14}} {S}^{14} {U}^{t_{14},t_{13}} {S}^{13} {U}^{t_{13},-R}.
$$
 Trivial phases resulting from the adiabatic evolution drop out from the matrix  of  transition probabilities   and we obtain  \cite{suppl}
\be
\label{prob1}
&&{ P}^{ v<b_1-b_2}=\left(
\begin{array}{cccc}
p_1p_2 & 0 & p_2q_1 & q_2\\
0 & p_1p_2 & q_2 & p_2 q_1 \\
p_2q_1 & q_2 & p_1p_2 & 0 \\
q_2& p_2q_1 & 0 & p_1p_2
\end{array}
\right),\\
\nonumber p_1 &=& e^{-2\pi g^2/(b_1-b_2)}, \,\,\, p_2=e^{-2\pi \gamma^2/(b_1+b_2)}, \,\,\, q_{1,2}=1-p_{1,2}.
\ee
This result does not depend on  $v$, so it coincides with the solution for the model  (\ref{ham1})   found in \cite{four-LZ,constraints}.

The situation changes for $b_2+b_2>v>b_1-b_2$. Now the points of adiabaticity violation $e_{24}$ and $e_{13}$  lie on the first leg of the path ${\cal P}_\infty$ as shown in Fig.~\ref{contour1}(b). Pairwise transitions near these points  are now governed by the Hamiltonian
${H}_1$ and  the transition probability matrix in this case  is different
\be
{P}^{ v>b_1-b_2}_{v<b_1+b_2}=\left(
\begin{array}{cccc}
p_1p_2 & q_1q_2 & p_2q_1 & p_1 q_2\\
q_1q_2 & p_1p_2 & p_1 q_2 & p_2 q_1 \\
p_2q_1 & p_1q_2 & p_1p_2 & q_1q_2 \\
p_1q_2& p_2q_1 & q_1q_2 & p_1p_2
\end{array}
\right).
\label{prob2}
\ee
For $v>b_1+b_2$, all  four points with Landau-Zener transitions   lie on the first leg of ${\cal P}_{\infty}$ and
\be
 {P}^{ v>b_1+b_2}=\left(
\begin{array}{cccc}
p_1p_2 & 0 &q_1 &  p_1q_2\\
0 & p_1p_2 & p_1q_2 &  q_1 \\
q_1 & p_1q_2 & p_1p_2 & 0 \\
p_1q_2& q_1 & 0 & p_1p_2
\end{array}
\right).
\label{prob3}
\ee
We see that our approach   not only reproduces the previously known solution for the Hamiltonian (\ref{ham1}), but also solves a more complex model (\ref{heff1}).



Thus, we have identified a symmetry -- multi-time evolution  with commuting Hamiltonians -- that  leads to the integrability of unitary dynamics with  time-dependent Hamiltonians. Our approach generates numerous new solvable multistate Landau-Zener models.
As  examples, we solved   a four-state  model   (\ref{heff1}) and proved the previously conjectured solution of a combinatorially complex driven Tavis-Cummings model \re{TC}, which has important applications in physics of molecular condensates  \cite{gurarie-LZ,altland3}. We believe, this symmetry is behind most if not all nontrivial exactly solvable  multistate Landau-Zener  and  Landau-Zener-Coulomb models \cite{sinitsyn-14pra,LZC,LZC1,LZC2}.  It explains why in such problems the scattering matrix factorizes into a product of two-state scattering matrices \cite{six-LZ} -- since Eq.~(\ref{system1}) allows a choice of an integration path that bypasses the region of complex nonadiabatic dynamics. It also explains why basic known solvable models have commuting partners with simple linear or quadratic dependence on $t$ \cite{yuzbashyan-LZ}. Indeed, pairs of such operators that also satisfy \eref{commute-p2}  lead to  relatively simple versions of the WKB approximation necessary to determine scattering matrices. Further,   Eq.~(\ref{comb1})  shows how certain  distortions of the parameters \cite{quest-LZ} give rise to entire families of solvable models. 

Finally, we note  that when $\hat{H}_j$ are isotropic Gaudin magnets [\eref{gaudin7} at $B=0$],    the $j=1,\dots,M$ subsystem of \eref{system1} is the famous Knizhnik-Zamolodchikov equation \cite{KZ}. Its solutions  have been obtained using off-shell Bethe's Ansatz \cite{babujian2}.  This   was generalized to  $B\ne 0$ in \cite{sedrakyan}  (see also \cite{sierra})    and exploited in \cite{gritsev} to obtain the dynamics of an isotropic Gaudin magnet with time-dependent $\eps_i$. We believe, solutions to \eref{system1}, i.e. exact inexplicit solutions of the non-stationary Shr\"odinger equation at arbitrary $t$, for all  time-dependent Hamiltonians discussed in this letter can be obtained by further extending this technique.

\acknowledgments
We thank A. Kamenev and S. L. Lukyanov for useful discussions. This work was carried out under the auspices of the National Nuclear Security Administration of the U.S. Department of Energy at Los Alamos National Laboratory under Contract No. DE-AC52-06NA25396 (N.A.S. and C.S.). It was also supported by NSF: DMR-1609829 (E.A.Y. and A.P.) and  Grant No. CHE-1111350 (V.Y.C.). N.A.S. also thanks the support from the LDRD program at LANL.

\newpage

\section*{Supplemental material for ``Integrable time-dependent Hamiltonians"}

\subsection{Multidimensional WKB approximation}
\label{sec:Int-LZ-general}

In the examples in the main text, we were able to obtain explicit transition probability matrices,  because the energy levels of Hamiltonians $\hat  H_{j} (\bm{x}) $ were  for the most part well separated   at large $|{\bm x}| $,  making the adiabatic approximation exact when $|{\bm x}|\to\infty $. In this section, we study this method generally, starting from the zero curvature condition for
real symmetric Hamiltonians, i.e. from Eqs.~(7,8) in the main text. We  interpret it as a multidimensional WKB method in the real  space $\mathbb{R}^{M+1}$. The elements of this space are
$\bm x=(t,x^1,\dots,x^M)$, where  $x^1,\dots,x^M$ are the system parameters and $t$ is the time variable.  In the ordinary WKB method, the  eigenstates  of a 1D or a multidimensional
completely separable Hamiltonian are proportional to $e^{i {\cal S}(\bm q, t)}$, where $\bm q$ are the generalized coordinates and ${\cal S}(\bm q, t)$ is the classical action. In our case, we cast the $\bm x$-dependence of the components of the wavefunction in the adiabatic basis  into the form $e^{i {\cal S}^{a}(\bm{x})}$, where $a=1,\ldots, N$ is the index of the adiabatic level.   The quantities ${\cal S}^{a}(\bm{x})$ are   single-valued functions
thanks to the path-independence of the time-ordered exponential
\begin{eqnarray}
 \label{time-ordsp}
\Psi(\bm{x}) = T \exp\left(-i \int_{{\cal P}} \hat{H}_{j} dx^{j}\right) \Psi(\bm{x_{0}}),  
\end{eqnarray}
[see Eq.~(3) in the main text] and they also turn out to be real. We therefore interpret
${\cal S}^{a}(\bm{x})$ as the classical action corresponding to the $a$-th adiabatic level. 

Eqs.~(7,8) from the main text read
\begin{eqnarray}
[\hat H_{j} (\bm{x}), \hat  H_{k}(\bm{x})] = 0,&\quad j,k=0, 1,\ldots, M. \label{curv-zero-real1} \\
\partial_{j} \hat  H_{k} (\bm{x}) - \partial_{k} \hat  H_{j} (\bm{x}) = 0. & \label{curv-zero-real2} 
\end{eqnarray}
 \eref{curv-zero-real1} implies that there is a basis    $|e_{a}(\bm{x}) \ra$ (adiabatic basis),  where $\hat H_{j} (\bm{x})$ are simultaneously diagonal, i.e. \begin{eqnarray}
\label{adiabatic-BS} H_{j}(\bm{x}) |e_{a}(\bm{x}) \ra&=& -p_{j}^{a}(\bm{x}) |e_{a}(\bm{x}) \ra,   \\ 
H_{j}(\bm{x}) &=& -\sum_{a} p_{j}^{a}(\bm{x}) | e_{a}(\bm{x})\rangle \langle e_{a}(\bm{x}) |, \label{adiabatic-BS1}
\end{eqnarray}
 where $-p_{j}^{a}(\bm{x})$ are the adiabatic levels. The substitution of  Eq.~(\ref{adiabatic-BS1}) into   Eq.~(\ref{curv-zero-real2})  yields a matrix equation, whose diagonal and off-diagonal  parts are
\begin{eqnarray}
\label{integr-adiabatic-BS} \partial_{j} p_{k}^{a}(\bm{x}) - \partial_{k} p_{j}^{a}(\bm{x}) &=& 0,   \\ 
\lambda_{j}^{ab}(\bm{x}) {\cal B}_{k}^{ab}(\bm{x}) - \lambda_{k}^{ab}(\bm{x}) {\cal B}_{j}^{ab}(\bm{x}) &=& 0,\label{integr-adiabatic-BS1}
\end{eqnarray}
  respectively. Here  
\begin{eqnarray}
\label{integr-adiabatic-BS-2} \lambda_{j}^{ab}(\bm{x}) &=& p_{j}^{a}(\bm{x}) - p_{j}^{b}(\bm{x}),   \\ 
{\cal B}_{j}^{ab}(\bm{x}) &=& \langle e_{a}(\bm{x}) | \partial_{j} e_{b}(\bm{x}) \rangle, \label{integr-adiabatic-BS-21} 
\end{eqnarray}
and ${\cal B}_{j}^{ab}(\bm{x})$  are known as the non-adiabatic coupling terms.

 Equation~(\ref{integr-adiabatic-BS}) implies $\bm{p}^{a}(\bm{x})$ are gradients, whereas \eref{integr-adiabatic-BS1} means that  vectors $\bm{{\cal B}}^{ab}(\bm{x})$ and $\bm{\lambda}^{ab}(\bm{x})$ are collinear,  i.e.
\begin{eqnarray}
\label{integr-adiabatic-BS-3} p_{j}^{a}(\bm{x}) &=& \partial_{j} {\cal S}^{a}(\bm{x}),  \\ 
{\cal B}_{j}^{ab}(\bm{x}) &=& \kappa^{ab}(\bm{x}) (p_{j}^{a}(\bm{x}) - p_{j}^{b}(\bm{x})). \label{integr-adiabatic-BS-31}
\end{eqnarray}
  Equation~(\ref{integr-adiabatic-BS-3}) allows us to interpret $\bm{p}^{a}$ as the classical momentum corresponding to the classical action ${\cal S}^{a}$ associated with the $a$-th adiabatic surface (level).  Quantities  $\kappa^{ab}(\bm{x})$ have the meaning of {\it position-dependent non-adiabaticity parameters}, so that the evolution is purely adiabatic when $\kappa^{ab}(\bm{x}) \to 0$ for all $a \ne b$. Indeed, substituting the wavefunction in the form
\beg
|\Psi (\bm{x})\rangle=\sum_a \Psi^a (\bm{x})   | e_{a}(\bm{x}) \rangle
\en
into the multi-time Shr\"odinger equations [Eq.~(1) in the main text] and  sending $\kappa^{ab}(\bm{x})$ to zero, we derive
\beg
\label{Sol-flat-conn-semiclass} \Psi (\bm{x}) = \sum_{a} e^{i({\cal S}^{a}(\bm{x}) - {\cal S}^{a}(\bm{x}_{0}))} \Psi^{a}(\bm{x_{0}}) | e_{a}(\bm{x}) \ra.  
\en
A key consequence of Eqs.~(\ref{curv-zero-real1},\ref{curv-zero-real2}) is that the non-adiabaticity parameters $\kappa^{ab}(\bm{x})$ are the same for all commuting Hamiltonians $\hat H_j(\bm x)$. In other words, the time evolution at any given point $\bm x_0$ is equally adiabatic for all paths ${\cal P}$ passing through that point. The condition $\kappa^{ab}(\bm{x}) \ll1$ for all $a\ne b$, therefore, defines \textit{adiabatic domains} in the space $\mathbb{R}^{M+1}$ of time and system parameters. In these domains, \eref{Sol-flat-conn-semiclass} provides an accurate WKB approximation to \eref{time-ordsp} becoming exact in the limit $\kappa^{ab}(\bm{x}) \to 0$.

\begin{figure}
\includegraphics[width=6cm]{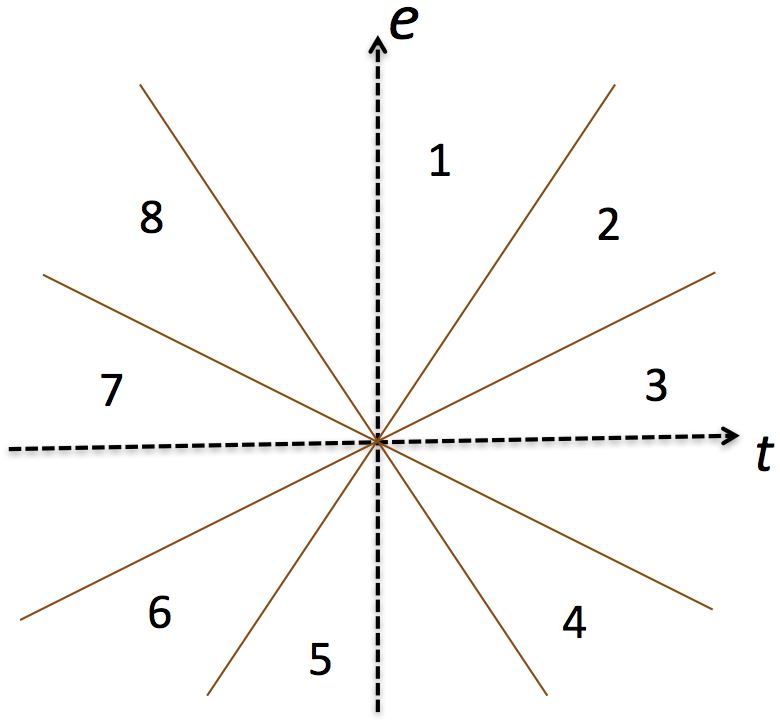}
\centering
\caption{(color online) Adiabatic domains for the 4-state problem. There are 8 domains separated by 4 solid lines $e=(\pm b_2\pm b_1)t$.  We determine these lines from the condition that a pair of diabatic energy levels of $H(t,e)$ in \eref{ham111} is degenerate, $E_i(t,e)=E_j(t,e)$. The time evolution is adiabatic at large $|\bm{x}|=\sqrt{t^2+e^2}$ in each domain away from its boundaries. WKB solutions in neighboring domains are connected via $2\times2$ Landau-Zener scattering matrices. Here $b_1=1$ and $b_2=0.5$.}
\label{domains-fig}
\end{figure}

Let us explicate this picture for the 4-state model analyzed in the main text. To determine   $\kappa^{ab}(\bm{x})$ and the adiabatic domains,  we can use any of the Hamiltonians, $H_0(t,e)\equiv H(t,e)$  or $H_1(t,e)$. Take
\be
{H} (t,e) = \left(
\begin{array}{cccc}
b_1t+e & 0 & g &-\gamma \\
0 & -b_1t +e & \gamma & g \\
g& \gamma & b_2 t & 0\\
-\gamma & g &0 &-b_2t
\end{array}
\right),
\label{ham111}
\ee
When $t$ and $e$ are large, the off-diagonal part of $H(t,e)$ is typically negligible. Then, the adiabatic and diabatic energy levels and eigenstates coincide. To the leading order, $\langle e_{a}(\bm{x}) | \partial_{t} e_{b}(\bm{x}) \rangle=0$, $p_{0}^{a}(\bm{x}) - p_{0}^{b}(\bm{x})\to\infty$ and
\beg
  \kappa^{ab}(\bm{x}) =\frac{\langle e_{a}(\bm{x}) | \partial_{t} e_{b}(\bm{x}) \rangle}{p_{0}^{a}(\bm{x}) - p_{0}^{b}(\bm{x})}=0,
\en  
i.e. the dynamics are purely adiabatic.

 Adiabaticity breaks down when two of the diabatic energies $E_1=b_1t+e, E_2=-b_1t+e, E_3=b_2t$ and $E_4=-b_2t$ are close. Consider, for example, levels
 $E_2$ and $E_3$. First order perturbation theory in the  off-diagonal part of $H(t,e)$ yields
 \beg
  \kappa^{23}(\bm{x}) = \frac{\gamma(b_1+b_2) }{(E_3-E_2)^3}=\frac{\gamma(b_1+b_2) }{[(b_1+b_2)t-e]^3}.
\en  
The breakdown occurs near the line $e=(b_1+b_2)t$ on which $E_2(t,e)=E_3(t,e)$. However, the nonadiabatic region where $\kappa^{23}(\bm{x}) \ge 1$ is confined inside a small angle of order $1/t$, whose bisector is the $e=(b_1+b_2)t$ line. Altogether equations $E_1(t,e)=E_3(t,e)$,
$E_1(t,e)=E_4(t,e)$, $E_2(t,e)=E_3(t,e)$ and $E_2(t,e)=E_4(t,e)$ define four lines $e=(\pm b_2\pm b_1)t$ in the coordinate space $(t,e)$. They divide the $(t,e)$ plane into eight adiabatic domains shown in Fig~\ref{domains-fig}. The remaining two degeneracies $E_1(t,e)=E_2(t,e)$ and $E_3(t,e)=E_4(t,e)$ are insignificant, because these levels are not directly coupled to each other (matrix elements $H_{12}$ and $H_{34}$ are zero).

An important ingredient of the ordinary WKB method in quantum mechanics is the matching conditions near the turning points.   In our multidimensional case,   the hypersurfaces, where $\kappa^{ab}$ is large and the semiclassical approximation  (\ref{Sol-flat-conn-semiclass}) breaks down, play the role of the turning points. In the 4-state example above these are the four lines where two of the diabatic levels are degenerate, see Fig~\ref{domains-fig}.
Similarly, in the driven generalized Tavis-Cummings model in the main text, the adiabaticity is violated when two levels get close. In cases like these, we obtain the matching (scattering) conditions by solving a 2$\times$2 scattering problem for these two states, say $a$ and $b$. The rest of the levels continue to evolve adiabatically, since they are well separated from levels $a$ and $b$ and from each other.
 Then, there is the following linear relationship between the semiclassical solutions   (\ref{Sol-flat-conn-semiclass}) in neighboring domains $\alpha$ and $\beta$ separated by a hypersurface on which levels $a$ and $b$ are degenerate: 
\begin{eqnarray}
\label{scatter-multidim} \Psi_{\alpha}^{c} = \sum_{d=a,b} S^{cd}_{\alpha\beta} \Psi_{\beta}^{d},\quad S_{\alpha\beta} = \bar{S}_{\alpha\beta; ab} \oplus \bar{I}_{ab}, 
\end{eqnarray}
 where $c$ takes values $a$ and $b$, $\bar{S}_{\alpha\beta; ab}$ is a $2 \times 2$ Landau-Zener scattering matrix  for states $a$ and $b$ evaluated near the degeneracy hypersurface, and  $\bar{I}_{ab}$ is  a unit matrix acting on the remaining states.      
   
 We see that the zero curvature condition makes an explicit solution of the scattering problem possible in two ways. First, it allows us to choose a path ${\cal P}_\infty$ connecting the initial and final points that goes through a series of adiabatic domains at large $|\bm{x}|$. Second, in each such domain it facilitates a multidimensional WKB approach. Relatively simple scattering matrices connect WKB solutions in neighboring domains. In our examples, these were $2 \times 2$  Landau-Zener  matrices, but other solvable systems may have other matching conditions. We thus obtain the desired scattering matrix for the evolution from $t=-\infty$ to $t=+\infty$ by going sequentially from the domain that contains  $t=-\infty$ to the one containing $t=+\infty$ always keeping $|\bm{x}|$ large and using Eqs.~(\ref{Sol-flat-conn-semiclass}) and \re{scatter-multidim} along the way. For example, to determine the transition probabilities for $H(t,e)$ in \eref{ham111}, we need to go from domain \#7 to domain \#3 in  Fig~\ref{domains-fig}.

\subsection{Transition Probabilities for the 4-State Model: Direct Calculation}
\label{sec:4-state-direct}

In this section, we  detail the calculation of  transition probabilities for the new 4-state model $h(\tau)$ [Eq.~(16) in the main text].
 In particular,  we show how the phases accumulated during adiabatic evolution and Landau-Zener scattering in between adiabatic domains drop out from the final result.  
 
 Consider the case $v<b_1-b_2$ and the path ${\cal P}_{\infty}$ shown in Fig.~2(a) of the main text. This path goes from domain \#7 to domain \#3 in  Fig~\ref{domains-fig}. Its horizontal part crosses the four lines in  Fig~\ref{domains-fig} where the adiabaticity is violated at points $t_{13}, t_{14}, t_{23}$ and $t_{24}$ marked with crosses in Fig.~2(a). The vertical piece of ${\cal P}_{\infty}$ does not contain points with nonadiabatic transitions. The evolution along this piece is adiabatic, described by \eref{Sol-flat-conn-semiclass}, and does not affect  the final transition probabilities.
 
\begin{figure}[t!]
\vskip +0.2in
\includegraphics[width=7.5cm]{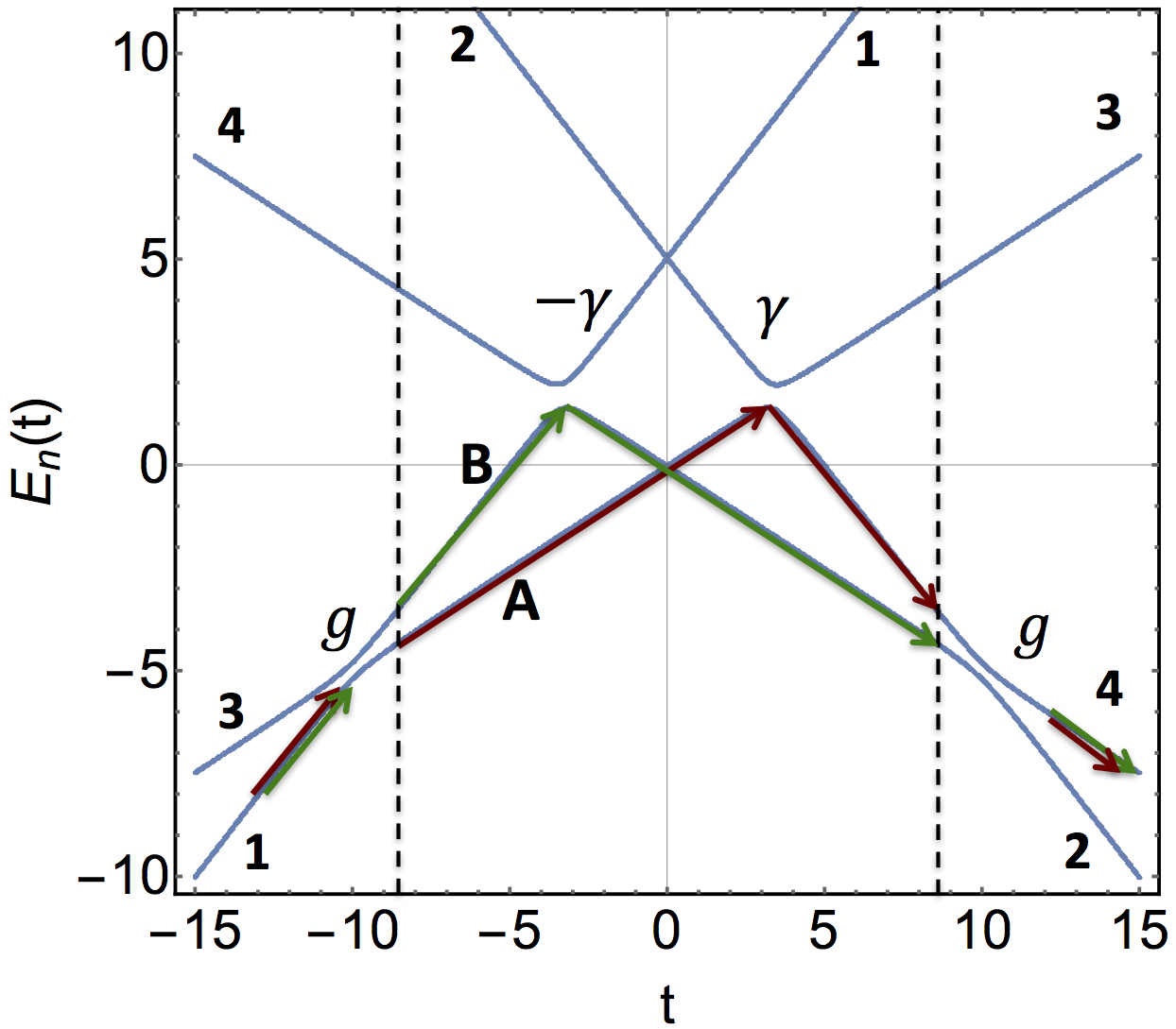} 
\centering
\caption{(color online)  Energies of the Hamiltonian (\ref{ham111})  as functions of $t$ (adiabatic levels). Numbers  1 through 4 indicate diabatic levels. For each avoided crossings  we display the corresponding pairwise coupling $g$ or $\pm\gamma$.
Red  and green arrows  show the two   trajectories $A$ and $B$ that contribute to the amplitude of the transition from diabatic level 1 to level 4. In the time interval between vertical dashed lines   these  
trajectories have different dynamical phases.  Here  $b_1=1, b_2=.5, e=5.0, g=0.2$, and $\gamma=0.3$. }
\label{spec-fig}
\end{figure}

 Therefore, it is sufficient to consider only the horizontal part of ${\cal P}_{\infty}$. Evolution along this fragment occurs with the  Hamiltonian \re{ham111},
 where $e=vR+e_0 \gg e_0,g,\gamma$. We show the adiabatic levels of this Hamiltonian in Fig.~\ref{spec-fig}. Due to the large value of $e$,  anticrossings are   well separated in energy from the rest of the levels.  
 The matrix of evolution along the horizontal piece of ${\cal P}_{\infty}$ is 
 \beg
\begin{split}
 {S} ={U}^{t=R,t_{24}} {S}^{24}_\mathrm{LZ} {U}^{t_{24},t_{23}}{S}^{23}_\mathrm{LZ}  {U}^{t_{23},t_{14}}\times\\
  {S}^{14}_\mathrm{LZ} {U}^{t_{14},t_{13}}
 {S}^{13}_\mathrm{LZ}{U}^{t_{13},t=-R},
\end{split}
\label{scatt}
\en
where $4\times4$ matrices ${U}^{\gamma,\delta}$ and   ${S}^{ij}_\mathrm{LZ}$ represent the adiabatic evolution between the time moments $\gamma$ and $\delta$ on the path ${\cal P}_{\infty}$ and pairwise Landau-Zener transitions between   states $i$ and $j$, respectively.  The  Landau-Zener amplitudes are known [see, e.g., Refs.~32-35 in the main text]. We have
\be
\nonumber {S}^{13}_\mathrm{LZ}=\left(
\begin{array}{cccc}
\sqrt{p_1} & 0 &i\sqrt{q_1} e^{i\phi_{g}} &0 \\
0 & 1 & 0 & 0 \\
i\sqrt{q_1} e^{-i\phi_g}  & 0 & \sqrt{p_1} & 0 \\
0&0&0& 1
\end{array}
\right),
\label{scatt13}
\ee
\be
{S}^{24}_\mathrm{LZ}=\left(
\nonumber \begin{array}{cccc}
1& 0 &0&0 \\
0 &\sqrt{p_1}& 0 & i\sqrt{q_1} e^{-i\phi_{g}} \\
0  &0& 1 & 0 \\
0& i\sqrt{q_1} e^{i\phi_g} &0& \sqrt{p_1}
\end{array}
\right),
\label{scatt24}
\ee
\be
\nonumber {S}^{14}_\mathrm{LZ}=\left(
\begin{array}{cccc}
\sqrt{p_2} & 0 &0 &-i\sqrt{q_2} e^{i\phi_{\gamma}} \\
0 & 1 & 0 & 0 \\
0  & 0 & 1 & 0 \\
-i\sqrt{q_2} e^{-i\phi_{\gamma}}&0&0& \sqrt{p_2}
\end{array}
\right),
\label{scatt14}
\ee
\be
\nonumber {S}^{23}_\mathrm{LZ}=\left(
\begin{array}{cccc}
1 & 0 &0 &0 \\
0 &\sqrt{p_2} &i\sqrt{q_2} e^{-i\phi_{\gamma}} & 0 \\
0  & i\sqrt{q_2} e^{i\phi_{\gamma}}  & \sqrt{p_2}& 0 \\
0&0&0& 1
\end{array}
\right).
\label{scatt15}
\ee
Here
$$
p_1 = e^{-2\pi g^2/(b_1-b_2)}, \,\,\, p_2=e^{-2\pi \gamma^2/(b_1+b_2)}, \,\,\, q_{1,2}=1-p_{1,2},
$$
and $\phi_g$, $\phi_{\gamma}$ are   transition phases associated with  couplings $g$ and $\gamma$ in \eref{ham111}, respectively. The sign in front of $\phi_{g,\gamma}$ depends on which level has a higher slope at the  crossing of the corresponding diabatic levels.
 Amplitudes associated with the couplings $- \gamma$  in \eref{ham111}  acquire additional minus signs in $S_\mathrm{LZ}^{14}$.

Let us, for example, evaluate the level 1 to level 4 transition amplitude $S_{1\rar 4}$, i.e. the matrix element $S_{14}$ of \eref{scatt}.  We work in the diabatic basis. As mentioned in the previous section,  far away from avoided crossings  diabatic    and adiabatic  bases coincide, i.e. matrices $U^{\gamma,\delta}$ in \eref{scatt} are diagonal. It follows from \eref{scatt} that there are two   ``trajectories" that contribute
to this amplitude -- shown with green and red arrows and marked as A and B in Fig.~\ref{spec-fig}, i.e.
\beg
S_{1\rar 4}=S^A+S^B.
\en
The adiabatic (dynamical) phases accumulated on these trajectories are 
\be
\phi_{d} = -\int_{-R}^{R} E (t) \, dt,
\label{dyn-ph}
\ee
which are the areas under the curves $A$ and $B$ in Fig.~\ref{spec-fig}.
 In order to make these phases well-defined, we  use a convention  that a trajectory  jumps from one adiabatic level  to another as the result Landau-Zener tunneling at the moment of the minimal   separation between the  levels. 
Due to the symmetry of the spectrum of the Hamiltonian (\ref{ham111}) under the reflection $t\rar -t$,   areas under the curves $A$ and $B$ are the same,
i.e. $\phi_d^A=\phi_d^B\equiv \phi_d$. This is true even though at intermediate times between vertical dashed lines in
Fig.~\ref{spec-fig}, the adiabatic phases of  trajectories $A$ and $B$ are different. 

Thus, using also the above expressions for the scattering matrices ${S}^{ij}_\mathrm{LZ}$, we  get
\begin{eqnarray}
S^{\rm A} &=& e^{i\phi_d}  (i\sqrt{q_1}e^{i\phi_g}) (i\sqrt{q_2} e^{-i\phi_{\gamma}}) (i\sqrt{q_1} e^{-i\phi_g}),\\
S^{\rm B} &=& e^{i\phi_d}  (\sqrt{p_1}) (-i\sqrt{q_2} e^{-i\phi_{\gamma}}) (\sqrt{p_1} ),
\label{tr-amp}
\end{eqnarray}
\be
S_{1\rar 4} = S^{\rm A} + S^{\rm B} =-i e^{i (\phi_d - \phi_{\gamma})} \sqrt{q_2},
\label{stot1}
\ee
and the corresponding transition probability,
\be
P_{1\rar 4} = |S_{1\rar 4}|^2 = q_2.
\label{ptot1}
\ee
We see that both  dynamic $\phi_d$ and  Landau-Zener $\phi_g$ and $\phi_{\gamma}$ phases  drop out from the final answer for the transition probability.   Note, however, that interference between the semiclassical trajectories $A$ and $B$ does take place despite this cancellation, $|S_{1\rar 4}|^2\ne |S^{\rm A}|^2 + |S^{\rm B}|^2$. For example, interference is responsible for the  independence of the final probability (\ref{ptot1})  from the coupling $g$, even though avoided crossings with this coupling  occur for both contributing  trajectories.

Similarly,   dynamical phases (\ref{dyn-ph}) as well as  $\phi_{\gamma}$ and $\phi_g$   cancel out  from all transition  probabilities due to the reflection symmetry in the spectra of   ${H}_{0}$ and ${H}_{1}$. Therefore, as long as we are interested in the transition probabilities only, we can disregard the matrices ${U}^{\alpha,\beta}$ in (\ref{scatt}) and set the Landau-Zener phases $\phi_{g,\gamma}$ in the scattering matrices ${S}^{ij}_\mathrm{LZ}$ to zero. The calculation of  ${S}$ then reduces to finding a properly ordered product of such truncated scattering matrices ${S}^{ij}_\mathrm{LZ}$.

\end{document}